\documentclass[aps,prl,twocolumn,superscriptaddress,showpacs]{revtex4}
\usepackage{amssymb}
\usepackage{graphics}
\usepackage{subfigure}
\usepackage{times}

\def\kfs122{KFe$_2$Se$_2$}
\def\kfsx22{K$_{1-x}$Fe$_2$Se$_2$}

\def\bfa122{BaFe$_2$As$_2$}
\def\fs11{FeSe}

\begin{document}

\preprint{February 20, 2011}
\title{Block spin magnetism and metal-insulator transition in a two-dimensional Hubbard model
with perfect vacancy superstructure}

\author{Hua Chen}
 % \email[E-mail address: ]{hchen@hznu.edu.cn}
  \affiliation{Zhejiang Institute of Modern Physics and Department of Physics,
  Zhejiang University, Hangzhou 310027, China}
\author{Chao Cao}
 % \email[E-mail address: ]{ccao@hznu.edu.cn}
 \affiliation{Condensed Matter Group,
  Department of Physics, Hangzhou Normal University, Hangzhou 310036, China}

\author{Jianhui Dai}
%\email[E-mail address: ]{daijh@zju.edu.cn}

\affiliation{Zhejiang Institute of Modern Physics and Department of
Physics, Zhejiang University, Hangzhou 310027, China}

\affiliation{Condensed Matter Group,
  Department of Physics, Hangzhou Normal University, Hangzhou 310036, China}

\date{February 22, 2011}

\begin{abstract}
We study the phase diagram of a square lattice Hubbard model with a
perfect vacancy superstructure. The model can be also defined on a
new bipartite lattice with each building blocks consisting of a
minimal square. The non-interacting model is exactly solved and a
mid-band gap opens at the Fermi energy in the weak inter-block
hopping regime. Increasing the Coulomb interaction will develop the
N\'eel antiferromagnetic order with varying block spin moments. The
metal-insulator transition with $U_{\rm MI}$ smaller than the one
without vacancies occurs above the magnetic instability $U_{\rm M}$.
The emergent intermediate magnetic metal phase develops
substantially in the moderate inter-block hopping regime. Drastic
increases in the ordered moment and the gap magnitude are observed
on the verge of tight-binding band insulator with increasing $U$.
The implications of these results for the recent discovered
(A,Tl)$_{y}$Fe$_{2-x}$Se$_2$ compounds are discussed.

\end{abstract}


\maketitle %\maketitle must follow title, authors, abstract and \pacs

The discovery of superconducting iron pnictide materials have
attracted enormous enthusiasm in searching for copper-free high
temperature superconductors\cite{lofa_discovery}. Recently, a family
of ternary iron chalcogenides (A,Tl)$_{y}$Fe$_{2-x}$Se$_2$
\cite{xlchen_prb_82_180520,mhfang_1012} (with $A$ being the alkali
atoms)has been shown to exhibit superconductivity (SC) in proximity
to a magnetic insulating phase. In the iron deficient compounds
($x>0$) there are Fe-vacancies in the square lattice of the
Fe-atoms. Interestingly, the Fe-vacancies could be ordered in
certain periodic superstructures in (K,Tl)Fe$_{2-x}$Se$_2$ where the
SC appears for $x=0.12 \sim 0.3$ \cite{mhfang_1012,arxiv:1101.0462}.
The vacancy superstructures were reported in early M\"{o}ssbauer
experiment on the non-superconducting material TlFe$_{2-x}$Se$_2$
some years ago\cite{Seidel}, and recent transmission electron
microscopy experiment \cite{arxiv:1101.2059} provides clear evidence
for the Fe-vacancy superstructures in the KFe$_{2-x}$Se$_2$ samples.

First principle calculations for (K,Tl)Fe$_{1.5}$Se$_2$ show that
the Fe-vacancy orthorhombic superstructure can be stabilized in the
ground states and play a crucial role in the electronic
structures\cite{arxiv:1012.6015,arxiv:1101.0533}. The observed
activation gap, which is about $\sim 60$ meV for $x\sim 0.5$, raises
a concern as whether the insulating behavior is due to the
antiferromagnetic (AFM) ordering itself, or due to the Fe-3d
electron correlations. The band structure calculations in
\cite{arxiv:1101.0533} provide evidence that a moderate Coulomb
interaction is required in order to account for the sizable gap and
a Mott localization can be enhanced due to the ordered vacancies.
The paramagnetic (PM) Mott transition for $x$ = 0.5 as well as
possible s-wave SC in a doped Mott-insulator have been also proposed
in a two-orbital model using spin-rotor mean-field (MF)
theory\cite{YuZhuSi, ZhouZhang}.

Here, we focus on another type of Fe-vacancy pattern, i.e., a
tetragonal vacancy superstructure shown in Fig. 1(a). The advantages
of this pattern is of two-folds. First, it is a perfect vacancy
superstructure in the sense that it exhibits the maximal symmetry
without breaking the in-plane four-fold rotational invariance and
that all sites are equally three-coordinated (with vacancy density
$20\%$)\cite{CaoDai}. Second, this structure has been clearly
observed in a number of (K,Tl)$_y$Fe$_{2-x}$Se$_2$ samples (for $x
\sim 0.4$)\cite{arxiv:1101.2059,arxiv:1101.4882,arxiv:1102.0830}.
Recent first principle calculations\cite{CaoDai} show that the
ground state of (K,Tl)$_y$Fe$_{1.6}$Se$_2$ is a large block spin
checkerboard or N\'eel AFM which is quite different to all
previously known magnetic patterns in the iron-based materials. This
novel AFM ordering pattern has been observed in recent neutron
diffraction
experiments\cite{arxiv:1102.0830,arxiv:1102.2882,arxiv:1102.3380}.
It is due to this large spin magnetism that the parent material
(K,Tl)$_{0.8}$Fe$_{1.6}$Se$_2$ could be an AFM
insulator\cite{CaoDai,LuXiang2}.

These developments motivate us to study a simplified single orbital
Hubbard lattice with the perfect vacancy superstructure. This model
not only describes the Mott physics in the presence of the ordered
vacancies, but also captures the geometry frustration as well as the
structure distortion. Therefore, it can serve as a basic model
system for the metal-insulator(MI) transitions and the complicated
magnetic properties in the presence of the vacancy orderings. In
this paper, we show that the interplay among the electron
correlation, the geometry frustration, and the structure distortion
in such a vacancy superstructure will result in a rich phase
diagram.
\begin{figure}[htp]
  \subfigure[ ]{
\scalebox{0.130}{\includegraphics{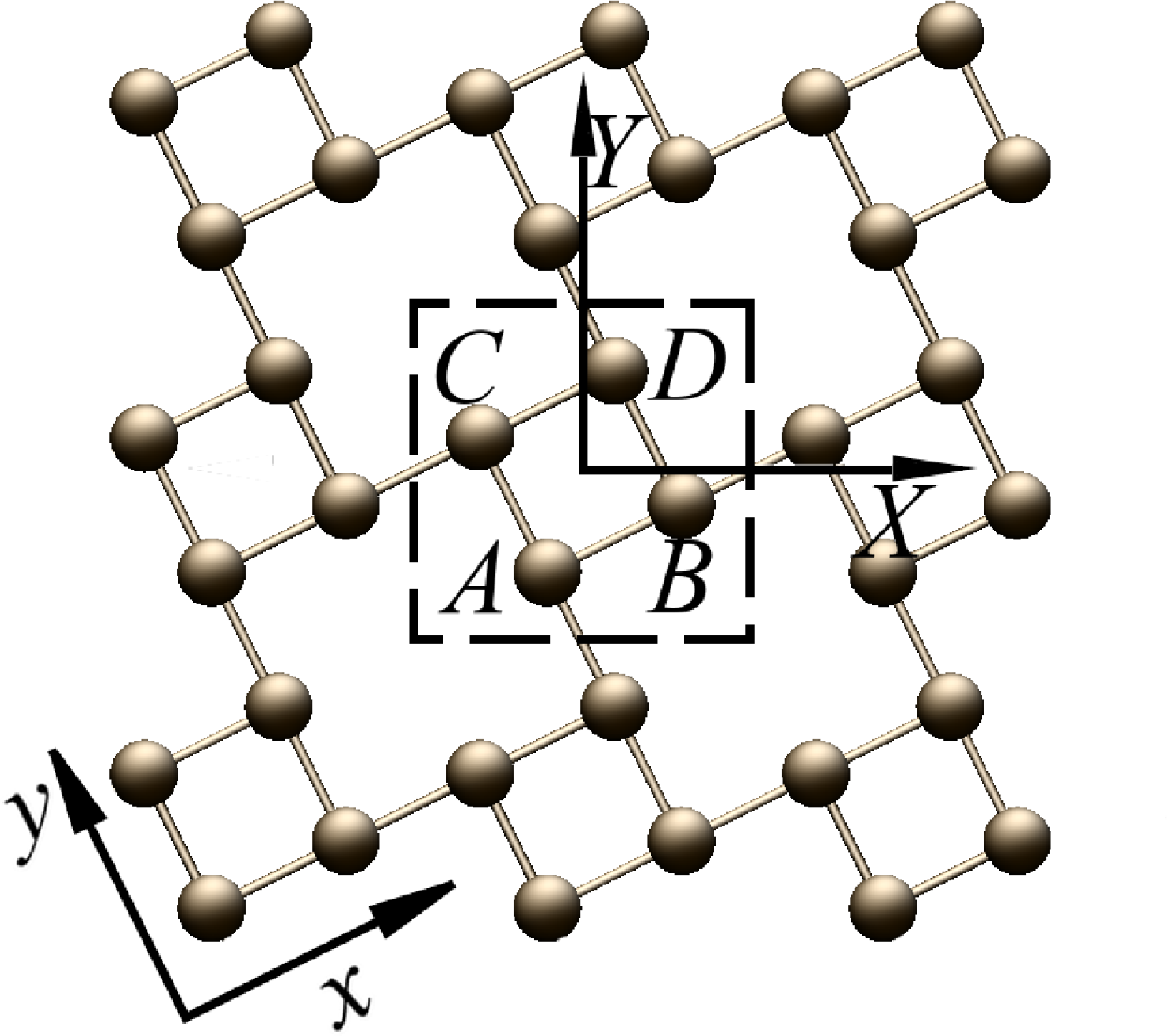}}
   \label{fig_superstructure}
  }
~~~\subfigure[ ]{
\scalebox{0.145}{\includegraphics{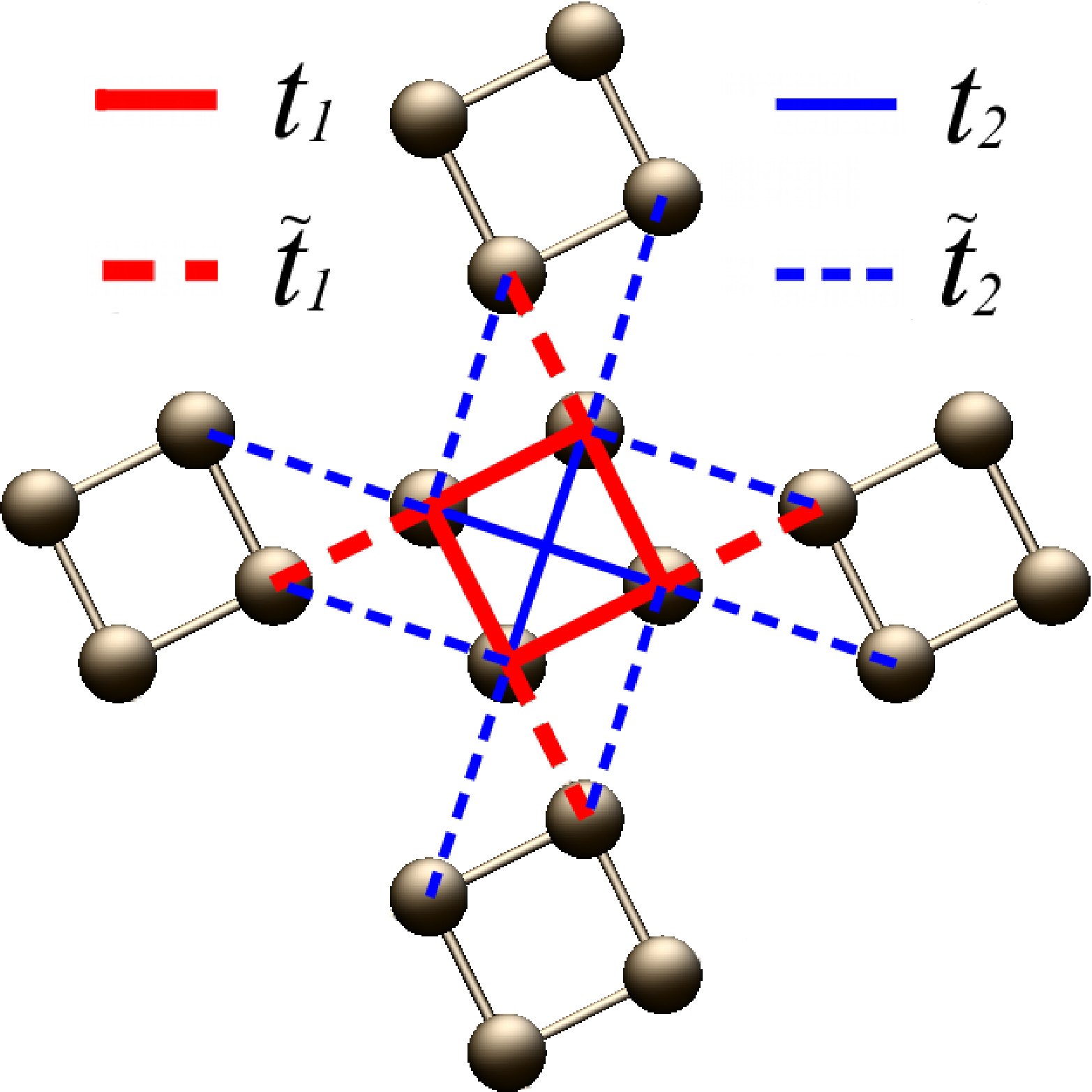}}
   \label{fig_interactions}
  }
 \caption{(color online). (a) Square lattice with perfect vacancy superstructure: The dashed
 lines connect four neighboring vacancies enclosing a minimal
 square. All sites are equivalent with three
 coordinates. The $x$, $y$ axes are for conventional Fe-Fe square
 lattice, while the $X$, $Y$ axes are for the bipartite lattice of blocks.
 (b) The n.n. (red) and n.n.n. (blue) intra-block (solid)
 and inter-block (dashed) hopping amplitudes.
   \label{fig_structure}}
\end{figure}

The Hamiltonian we studied is the Hubbard model
\begin{eqnarray}
H=-\sum_{ij\sigma} t_{ij} c^{\dagger}_{i\sigma}c_{j\sigma} +
\sum_{i}Un_{i\uparrow}n_{j\downarrow}+\sum_{i}\mu
(n_{i\uparrow}+n_{i\downarrow}),
 \label{hamiltonian1}
\end{eqnarray}
where $c_{i\sigma}$ is the annihilation operator for a Fe-3d
electron at site $i$ with spin $\sigma=\uparrow,\downarrow$;
$n_{i\sigma}=c^{\dagger}_{i\sigma}c_{i\sigma}$ the electron number
operator; $t_{ij}$ the hopping parameters; $U$ the on-site Coulomb
interaction; and $\mu$ the chemical potential. As a peculiar feature
of the vacancy superstructure, the whole lattice can be perfectly
covered by fundamental square blocks enclosed by four neighboring
vacancies under the periodic boundary condition, see in Fig.1(a),
where the $X$, $Y$ are the new axes for the block lattice. Then the
four sites in each block can be labeled by $A=(5m-1/2, 5n-3/2)/\sqrt
5$, $B=(5m+3/2, 5n-1/2)/\sqrt 5$, $C=(5m-3/2, 5n+1/2)/\sqrt 5$, and
$D=(5m+1/2, 5n+3/2)/\sqrt 5$, where, integers $(m,n)$ labels the
block position and the shortest Fe-Fe distance is set to be unit.
Hence the model is invariant under translations $m\rightarrow m+1$
or $n\rightarrow n+1$  along either directions. We then let
$t_{i,j}=t_1,t_2$ ( or $t_{i,j}={\tilde t}_1,{\tilde t}_2$ ) be the
intra-block ( or inter-block ) nearest neighbor (n.n.) and the next
nearest neighbor (n.n.n.) hopping amplitudes, illustrated in
Fig.1(b). Notice that the n.n.n. hopping amplitudes are mainly due
to the 3d-4p hybridization bridged by the Se-atoms above or below
each blocks(not shown in Fig.1(a)), and in general, the vacancies
will lead to structure distortions
\cite{arxiv:1101.0533,CaoDai,LuXiang2}, breaking the degeneracy
between the intra- and inter-block hopping amplitudes. While these
hopping parameters can be determined by fitting the LDA band
structures, our task here is to study the phase diagram for generic
model parameters.

To begin with, we first exactly solve the non-interacting
tight-binding model, $H_0=\sum_{{\vec
K}\lambda\sigma}c^{\dagger}_{\lambda\sigma}({\vec
K})H_{\lambda\lambda'}({\vec K})c_{\lambda\sigma}({\vec K})$, where,
$c_{\lambda\sigma}({\vec K})$ is the Fourie transformation of
$c_{\lambda\sigma}(p)$ for an electron at the site $\lambda$ in the
$p$-th block, ${\vec K}$ is in the first Brillouin zone (BZ) of the
periodic blocks, $\lambda,\lambda'=A,B,C$ and $D$, and
$H_{\lambda\lambda'}=[0, t_1+{\tilde t}_2e^{-iK_x}, t_1+{\tilde t}_2
e^{-iK_y}, t_2+{\tilde t}_1 e^{-iK_y}]; [t_1+{\tilde t}_2
e^{iK_x},0,t_2+{\tilde t}_1 e^{iK_x},t_1+{\tilde t}_2 e^{-iK_y}];
[t_1+{\tilde t}_2 e^{iK_y},t_2+{\tilde t}_1 e^{-iK_x},0,t_1+{\tilde
t}_2 e^{-iK_x}];[t_2+{\tilde t}_1 e^{iK_y},t_1+{\tilde t}_2
e^{iK_y},t_1+{\tilde t}_2 e^{iK_x},0]$. For simplicity, we assume
$t_2/t_1={\tilde t}_2/{\tilde t}_1=\alpha$, ${\tilde
t}_1/t_1={\tilde t}_2/t_2=\beta$ in the following discussions
(setting $t_1=1$), the extensions to more generic cases are
straightforward. Hence, $\alpha$, $\beta$ measure the strengths of
hopping frustration (due to the Se-bridged hybridization) and the
structure distortion (due to the Fe-vacancies) respectively. The
obtained four bands $\varepsilon_{1,2,3,4}({\vec K})$ are
illustrated via the band structures and the density of state (DOS)
for several sets of ($\alpha$, $\beta$), shown in Fig. 2(a-d). The
sub-band gaps $\Delta_{12}$,$\Delta_{34}$ as well as the mid-band
gap $\Delta_{23}$ are plotted in Fig.3(a-c). We find that the system
is always metallic at half-filling except for $\beta\lesssim 0.18$,
where the mid-band gap $\Delta_{23}$ opens at the Fermi energy for
$\alpha \gtrsim 1$.
\begin{figure}[htp]
  \centering
    \scalebox{0.35}{\includegraphics{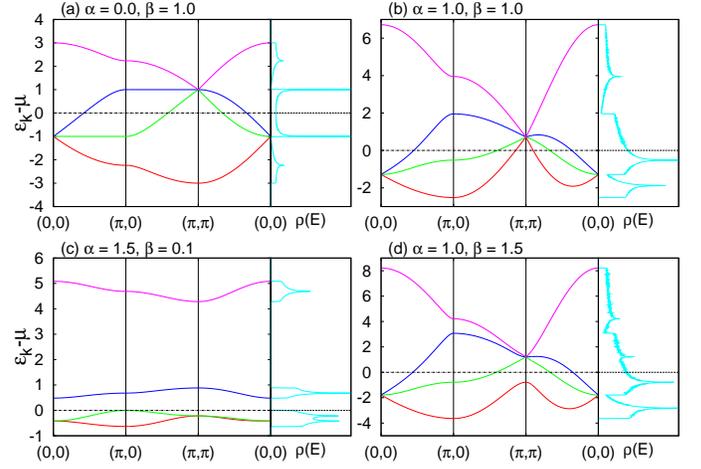}}
    \label{fig_gap}
\caption{(color online). Non-interacting band structures and DOS
$\rho(E)$. }
\end{figure}
\begin{figure}[htp]
  \centering
\scalebox{0.38}{\includegraphics{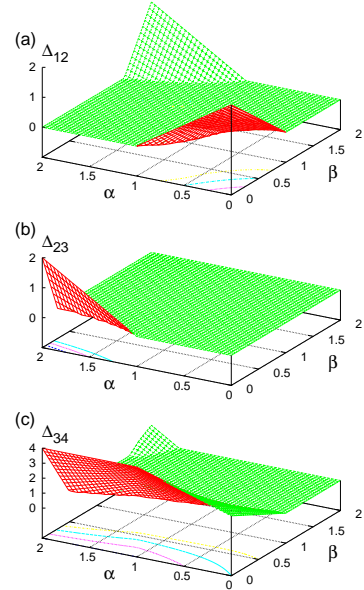}} \label{fig_gap}
  \caption{(color online). Various non-interacting band gaps : (a) sub-band gap $\Delta_{12}$;
  (b) mid-band gap $\Delta_{23}$;
  (c) sub-band gap $\Delta_{34}$. }
\end{figure}

On the other hand, in the large $U$ limit, the n.n. and n.n.n.
hoppings lead to the AFM exchanges ($J_1=4t^2_1/U$, $J_2=4t^2_2/U$)
and (${\tilde J}_1=4{\tilde t}^2_1/U$, ${\tilde J}_2=4{\tilde
t}^2_2/U$) for the intra- and inter-block spins, respectively. The
resulting spin Hamiltonian reads
\begin{eqnarray}
H=\sum_{p}H_{p}+\sum_{p,q}H_{p,q},
\end{eqnarray}
where
\begin{eqnarray*}
H_{p}=\{\sum_{\lambda,\lambda'\in n.n.}J_1+\sum_{\lambda,\lambda'\in
n.n.n.}J_2\} ~{\vec S}_{\lambda}(p)\cdot {\vec S}_{\lambda'}(p)
\\
H_{p,q}=\{\sum_{\lambda,\lambda' \in n.n.}\tilde
{J}_1+\sum_{\lambda,\lambda'\in n.n.n.} \tilde{J}_2\}~{\vec
S}_{\lambda}(p)\cdot {\vec S}_{\lambda'}(q)
\end{eqnarray*}
are the intra- and inter-block Hamiltonians, respectively, with
$p,q=(m,n)$ being the n.n. blocks. The spin Hamiltonian is well
defined on the new bipartite lattice with AFM exchanges connecting
the even and odd sublattices. It can be further shown that the
lowest energy state has positive definite coefficients in the
rotated Ising basis (obeying the Marshall's sign
rule\cite{LiebMattis,note1}), hence the ground state is a singlet of
the total spin $\sum_{p}{\vec S}_{block}(p)$, where ${\vec
S}_{block}(p)=\sum_{\lambda}{\vec S}_{\lambda}(p)$ is the total spin
in the $p$-th block. This implies that for each fixed
$\lambda$-site, $\{{\vec S}_{\lambda}(p),\forall p \}$ forms a spin
lattice of period ${\sqrt 5}\times {\sqrt 5}$, resulting in the
N\'eel AFM orders in the classical limit [except for ${\vec
S}_{block}(p)=0$ where the ferromagnetic order is possible
\cite{note1}], while the relative directions among different spins
$\{{\vec S}_{\lambda}(p), \forall\lambda\}$ in a given block are
dependent on the intra- and inter-block couplings. Notice that the
block spin ${\vec S}_{block}(p)$ should not be understood as a
single entity, since the interaction cannot be rewritten in terms of
${\vec S}_{block}(p)$. Rather, its formation is protected by the
periodicity of the AFM bipartite lattice and a strong intra-block
coupling is not a prerequisite. The expectation value of the block
spin $\langle S^{z}_{block}(p)\rangle$ varies in between $0$ and $2$
and could reach the maximal magnitude when $2{\tilde J}_{2}+{\tilde
J}_{1}\geq 2J_{1}+J_{2}$.

We now consider the weak and intermediate $U$ and solve the model by
the MF method as applied to the conventional $t_1$-$t_2$ Hubbard
model\cite{PRB35,JPSJ65,Vollhardt98,Yin10}. We introduce the
magnetic order parameters $M_{\lambda}$ by writing
\begin{eqnarray}
\langle 2S^z_{\lambda}(p)\rangle \equiv \langle
n_{\lambda\uparrow}(p)\rangle-\langle
n_{\lambda\downarrow}(p)\rangle =M_{\lambda}\cos ({\vec Q}\cdot
{\vec r}(p))
\end{eqnarray}
for each $\lambda$-sites, with ${\vec r}(p)=(m,n)$ being the
position of the $p$-th block.
%The ordered moment of a block is then
%given by $M/2=\sum_{\lambda}M_{\lambda}/2$.
%Because there is no direct coupling between the n.n.n. blocks,
We also consider the wavevector ${\vec Q}=(\pi,\pi)$, i.e., the
N\'eel AFM ordering, for there is no direct coupling between the
n.n.n. blocks. In our model, the collinear AFM ordering ${\vec
Q}=(\pi,0)$ (or ${\vec Q}=(0,\pi)$) may be favored when $\langle
s^z_{block}(p)\rangle=0$, and this case is simply identified with
the collinear AFM in the $t_2$-$t_2$ Hubbard ( or $J_1$-$J_2$
Heisenberg ) model (without vacancies) favored for strong
frustration $t_2$ (or $J_2$)\cite{Yin10,Coleman90}.
%The situation is different to the $t_1$-$t_2$ Hubbard ( or
%$J_1$-$J_2$ Heisenberg ) model defined on the conventional square
%lattice where the collinear AFM state is favored for strong
%frustration $t_2$ (or $J_2$)\cite{Yin10,Coleman90}. Here, the
%difference is due to the vacancy symmetry.

The MF approximation for the on-site repulsion $U$-term is standard,
and diagonalization of the MF Hamiltonian is implemented over the
first magnetic BZ zone of $K$-space (the reduced BZ  for one of the
sublattice). Eight bands of quasi-particles $E_{1,2,\cdots,8}({\vec
K})$ are obtained, and the magnetic moments $M_{\lambda}$ as
functions of $U$ are self-consistently determined by tuning the
chemical potential $\mu$ at the half-filling.
%As the model is symmetric under
%$A\leftrightarrow D$ and $C\leftrightarrow B)$, we find
%$m_A+m_D=m_B+m_C$ in our MF calculations.
We first focus on the moderate frustration regime ($\alpha=0.8$)
with weak, moderate, and strong inter-block hopping amplitudes
exemplified with (a)$\beta=0.2$, (b) $\beta=1$, and (c) $\beta=1.5$,
respectively. For each case, the magnetizations $M_{\lambda}$ as
well as the magnitude of the mid-band gap $\Delta$ at the Fermi
energy as functions of $U$ are shown in Fig. 4(a-c). We find that
the four $M_{\lambda=A,B,C,D}$ are equal to each other in the
calculated cases whereas they could be different in more generic
cases. To draw the phase diagram, we perform the same calculations
by varying the frustration $\alpha\sim 0-1.5$ for each fixed
$\beta$. The AFM instability $U_{\rm M}$ and the MI transition point
$U_{\rm MI}$ are determined by the onsets of magnetization and gap
within the accuracy $\sim 10^{-6}$. For comparison, another point
$U_{\rm vH}$ is marked at the kink position in the magnetization
curves in Fig.4. This point is identified as the van Hove
singularity at the Fermi surface as evidenced by the corresponding
band structures(not shown). Then, the $\alpha-U$ phase diagrams are
plotted in Fig.5 for each cases. Notice that in Fig.5(c) the AFM
metal region above the dashed line ($\alpha\lesssim 0.8$ ) may
actually merge into the PM phase because the magnetization is
vanishingly small within the calculation accuracy. Finally, three
main aspects of these results are briefly discussed as follows.
\begin{figure*}[htp]
\scalebox{0.450}{\includegraphics{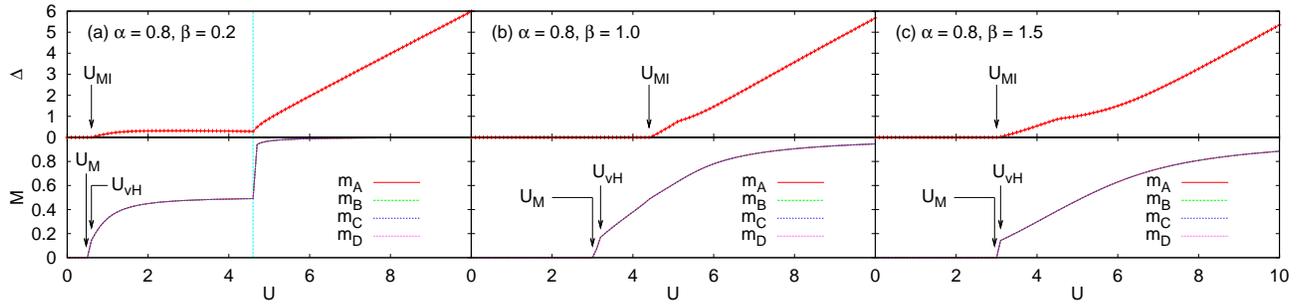}}
   \label{fig_maggap}
\caption{(color online).
 $U$-dependences of magnetic moments $M_{\lambda}$
 and the mid-band gap $\Delta$.
   }
\end{figure*}

\begin{figure*}[htp]
\scalebox{0.450}{\includegraphics{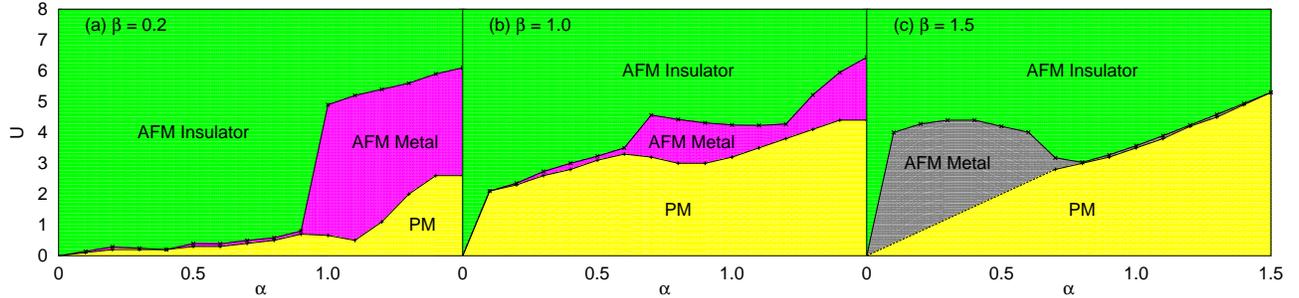}}
   \label{fig_phase}
\caption{
 (color online). Phase diagrams. The yellow, red, and green regions
 indicate the PM metal, AFM metal, and AFM insulator phases,
 respectively. Notice that in (c) the magnetization in the AFM metal region
 (pink region(light gray)) above the dashed line for $\alpha\lesssim 0.8$ is vanishingly
 small so that this AFM metal region may actually merge into the PM region
 (yellow region).
   }
\end{figure*}

(I) For moderate inter-block hopping ($\beta\sim 1$), the ground
state is a PM metal for sufficiently small $U$ and an AFM insulator
for relatively large $U$. An extended intermediate AFM metallic
phase emerges, separating the PM metal and AFM insulator phases in a
wide range of frustration $\alpha\sim 0.5-1.5$. For instance, when
$\alpha=0.8$, we find $U_{\rm MI}\sim 4.4$, and $U_{\rm M}\sim 3.0$,
respectively, while the van Hove singularity, which is located in
between $U_{\rm M}$ and $U_{\rm MI}$, could be clearly identified.
It is worth noting that in the corresponding $t_1$-$t_2$ Hubbard
model without vacancies the N\'eel AFM metal phase is extremely
narrow and is replaced by the collinear AFM metallic phase for
$\alpha\gtrsim 0.7$\cite{Yin10}. The corresponding $U_{\rm MI}\sim
5.0$, $U_{\rm M}\sim 3.6$. So we readily find that the Coulomb $U$
needed for the Mott-transition is lowered from $U_{\rm MI}\sim 5.0$
to $U_{\rm MI}\sim 4.4$. A similar conclusion for $U_{\rm MI}$ is
obtained by the spin-rotor MF theory for the two-orbital model
appropriate for the iron-deficient ternary iron
chalcogenides\cite{YuZhuSi}.

(II) In the strong inter-block hopping regime $\beta\sim 1.5$, the
distinction between the PM and AFM metal phases is not significant,
in particular for $\alpha\lesssim 0.8$, where the magnetization is
vanishingly small. Thus in this regime, the AFM ordering immediately
leads to the MI transition, and the van Hove singularity $U_{\rm
vH}$ is also very close to $U_{\rm MI}$. Another feature is that the
$U_{\rm MI}$ shows a minimum in the moderate frustration regime
around $\alpha\sim 0.8$, while the gap dependence on $U$ is almost
linear after the minimum. This feature resembles to that in the
moderate inter-block hopping regime, see in Fig.4(b-c).
%The common
%tendency of the linear gap dependence on $U$ in both cases is
%indicative of a Mott-insulator.

(III) When the non-interacting system is close to the band insulator
phase, as in the case of weak inter-block hopping $\beta\sim 0.2$
and for $\alpha\lesssim 0.8$, we find that the band-gap opens at the
Fermi energy immediately with very small $U_{\rm MI}$, after the
development of the N\'eel AFM order. The magnetization $M_{\lambda}$
rapidly saturates to $1/2$, and the gap magnitude $\Delta$, while
being very small, keeps almost unchanged for further increase of $U$
below a threshold value $U\sim 4.5$, manifesting itself as a
conventional AFM semi-conductor.  However, a drastic increase in
$M_{\lambda}$ is observed above this point, where it rapidly
saturates further to the maximal value (defined as unit), see in
Fig.4(a). Meanwhile, the gap increases almost linearly with the same
tendency seen in (I) and (II). Moreover, the PM phase shrinks and
$U_{\rm MI}$ increases for $\alpha\gtrsim 0.8$, so an intermediate
AFM metal phase develops substantially in the strong frustration
regime.

To summarize, we study the single-orbital Hubbard model on the
square lattice with the tetragonal vacancy superstructure, the most
perfect vacancy ordering pattern observed in the recently discovered
iron deficient compounds (A,Tl)$_{y}$Fe$_{2-x}$Se$_2$. With
increasing Coulomb interaction $U$ the model shows a block-spin type
N\'eel AFM order as a consequence of the vacancy symmetry. The MI
transition takes place at a relatively small $U_{\rm MI}$ compared
to the conventional Hubbard model without vacancies. But in a wide
range of parameters an extended intermediate region of the metal
phase with block-spin N\'eel AFM exists, separating the PM metal and
AFM insulator phases. This feature is also in contrast to the models
without vacancies. In some cases, an almost linear gap dependence on
$U$ accompanied by a broad crossover of the magnetization is
observed slightly above $U_{\rm MI}$ and this behavior is indicative
of the electron correlation effect within our MF approximation. An
interesting exception is the case for the weak inter-block hopping
regime, where drastic increases in both magnetization and gap
magnitude are observed in the insulator phase.
% revealing a clear distinction
%between the AFM semi-conductor and the AFM Mott-insulator.
Previous LDA calculations suggest that inter-block n.n.n. AFM
exchange ${\tilde J}_2$ dominates in the ground state of
(A,Tl)$_{0.8}$Fe$_{1.6}$Se$_2$\cite{CaoDai}, indicating that the
system should be in the strong inter-block hopping regime. While the
multi-orbital characteristic and the Hund's rule coupling should be
also taken into account for realistic materials, our results shed
new light in understanding the electron correlation and the novel
magnetism
%and possible Mott physics
in the (A,Tl)$_{y}$Fe$_{2-x}$Se$_2$ compounds.

%Previous LDA calculations for ( find that the ground state of the
%parent compound ($y=0.8$) is a block-spin N\'eel AFM insulator and the
%n.n.n. inter-block coupling ${\tilde J}_2$ dominates over other spin
%couplings\cite{CaoDai}. It implies that in this real five Fe-3d
%orbitals system, $\beta\sim 1$ or larger, i.e., it is in the strong
%inter-block hopping regime.
%Meanwhile, the recent neutron experiments observe a large ordered
%moment in the AFM phase\cite{arxiv:1102.0830,arxiv:1102.3456}.
%Based on the results obtained here, we suggest that the pressure
%dependence of the gap magnitude could be used to .

This work was supported by the NSFC, the NSF of Zhejiang Province
(Grant No. Z6110033), the 973 Project of the MOST, and the PCSIRT of
China ( Grant No. IRT-0754).

\bibliographystyle{apsrev}
\bibliography{Magnetic-Mott}

\begin{thebibliography}{23}
\bibitem[{not()}]{note1}
\bibinfo{note}{Violation of the sign rule may take place for $S^z_{block}(p)=1$
  ( or 0) if the breakdown of the $C_4$-rotational symmetry is allowed where
  permutations of $S^z_{\lambda}(p)$ among various $\lambda$ for fixed $p$ are
  possible.}
\end{thebibliography}
\end{document}